\documentclass[12pt]{article}
\usepackage{amsfonts}
\usepackage{amsmath}
\usepackage{amssymb}
\begin{document}

\title{Comment on \textquotedblleft Wave functions for a Duffin-Kemmer-Petiau particle in a time-dependent potential\textquotedblright}
\author{L.B. Castro\thanks{%
benito@feg.unesp.br } and A.S. de Castro\thanks{%
castro@pq.cnpq.br} \\
\\
UNESP - Campus de Guaratinguet\'{a}\\
Departamento de F\'{\i}sica e Qu\'{\i}mica\\
12516-410 Guaratinguet\'{a} SP - Brazil}
\date{}
\maketitle

\begin{abstract}
It is shown that the paper \textquotedblleft Wave functions for a Duffin-Kemmer-Petiau particle in a time-dependent potential\textquotedblright\, by Merad and Bensaid [J. Math. Phys. \textbf{48}, 073515 (2007)] is not correct in using inadvertently a non-Hermitian Hamiltonian in a formalism that does require Hermitian Hamiltonians.

\bigskip \bigskip \bigskip


\bigskip

\end{abstract}

In a recent paper published in this journal, Merad and Bensaid \cite{merad} reported on the solution of the Duffin-Kemmer-Petiau (DKP) equation for spin-0 sector in a time-dependent linear scalar field using the Lewis-Riesenfeld (LR) invariant method \cite{lewis}. The purpose of this comment is to point out that this alternative method is not applicable to the DPK theory.

The concept of invariance of a system, as introduced in Ref.\cite{lewis}, states that $I$ is an invariant of the system ($\hbar=c=1$) if
\begin{equation}\label{invariant}
    \frac{dI(t)}{dt}=\frac{1}{i}\left[ I(t),H(t) \right]+\frac{\partial I(t)}{\partial t}=0
\end{equation}

\noindent where $i$ is the imaginary unit, $I(t)$ and $H(t)$ (Hamiltonian of the system) are explicitly time-dependent operators. In this context, $I(t)$ and $H(t)$ should be Hermitians. On the other hand, the Hamiltonian form of the DKP particle with an electromagnetic interaction is written as
\begin{equation}\label{hf}
    H_{\mathrm{DKP}}=i\left[ \beta^{k},\beta^{0} \right]D_{k}+\frac{ie}{2m}\left( \beta^{\mu}\beta^{0}\beta^{\alpha}
    +\beta^{\mu}g^{0\alpha} \right)F_{\alpha\mu}+m\beta^{0}-eA^{0}
\end{equation}

\noindent where $D_{k}=\partial_{k}+ieA_{k}$ and $H_{\mathrm{DKP}}$ satisfies an equation of Schr\"{o}dinger type,
\begin{equation}\label{st}
    i\partial_{t}\psi=H_{\mathrm{DKP}}(t)\psi
\end{equation}

\noindent At this point, it is worthwhile to mention that $H_{\mathrm{DKP}}$ is not Hermitian \cite{tati}, as opposed to what was adverted in Ref.\cite{now}, since
\begin{equation}
\left( iF_{0i}\beta ^{i}\left( \beta ^{0}\right) ^{2}\right) ^{\dagger
}=-\left( iF_{0i}\beta ^{i}\left( \beta ^{0}\right) ^{2}\right)
+iF_{0i}\beta ^{i}  \label{her}
\end{equation}%
This results in
\begin{equation}
H_{\mathrm{DKP}}-H_{\mathrm{DKP}}^{\dagger }=\frac{ie}{m}F_{0i}\left[ \beta ^{i},\left( \beta
^{0}\right) ^{2}\right]\label{source1}
\end{equation}%

Therefore, it is not correct to use the LR invariant method in the DKP theory. In fact, the LR invariant theory generalized in Ref.\cite{gao} for including non-Hermitian Hamiltonians in the context of the nonrelativistic quantum mechanics might be appropriate to deal with the DKP Hamiltonian.

\bigskip

\noindent \textbf{Acknowledgments}

This work was supported in part by means of funds provided by CAPES and CNPq. The authors would like to thank the referees for drawing attention to Ref.\cite{gao}.

\end{document}